\documentclass[final,5p,times,twocolumn]{elsarticle}
\usepackage{graphicx}
\usepackage{amssymb}
\usepackage{lineno}
\usepackage{amssymb}
\usepackage[colorlinks,linkcolor=SB,citecolor=red]{hyperref}
\usepackage{geometry}
\usepackage{graphicx}
\usepackage{natbib}
\usepackage{etoolbox}
\usepackage{float}

\apptocmd{\sloppy}{\hbadness 10000\relax}{}{}

\biboptions{comma,sort&compress}
\journal{}

\begin{document}

\begin{frontmatter}


\title{Analysis of the Global Banking Network by Random Matrix Theory}



\author[namakiaddress1,namakiaddress2,ccnsdaddress]{Ali Namaki}
\ead{alinamaki@ut.ac.ir}
\author[ardalankiaaddress1,ccnsdaddress]{Jamshid Ardalankia}
\author[namakiaddress1]{Reza Raei}
\author[ccnsdaddress,Leilaaddress1]{Leila Hedayatifar}
\author[ccnsdaddress,grjafariaddress1]{Ali Hosseiny}
\author[Havenaddress]{Emmanuel Haven}
\ead{ehaven@mun.ca}
\author[ccnsdaddress,grjafariaddress1,grjafariaddress3]{G.Reza Jafari}
\ead{gjafari@gmail.com}

\address[namakiaddress1]{Department of Finance, University of Tehran, Tehran, Iran}
\address[namakiaddress2]{Iran Finance Association, Tehran, Iran}
\address[ardalankiaaddress1]{Department of Financial Management, Shahid Beheshti University, G.C., Evin, Tehran, 19839, Iran}
\address[ccnsdaddress]{Center for Complex Networks and Social Datascience, Department of Physics, Shahid Beheshti University, G.C., Evin, Tehran, 19839, Iran}
\address[Leilaaddress1]{New England Complex Systems Institute, NECSI HQ 277 Broadway, Cambridge, MA --- 02139, United States}
\address[Havenaddress]{Faculty of Business Administration, Memorial University, St. John's, Canada and IQSCS, UK}
\address[grjafariaddress1]{Department of Physics, Shahid Beheshti University, G.C., Evin, Tehran, 19839, Iran}
\address[grjafariaddress3]{Department of Network and Data Science, Central European University, 1051 Budapest, Hungary}

\begin{abstract}
Since 2008, the network analysis of financial systems is one of the most important subjects in economics. 
In this paper, we have used the complexity approach and Random Matrix Theory (RMT) for analyzing 
the global banking network. 
By applying this method on a cross border lending network, it is shown that the network has been denser
and the connectivity between peripheral nodes and  the central section has risen. 
Also, by considering the collective behavior of the system and comparing it with the shuffled one, 
we can see that this network obtains a specific structure. 
By using the inverse participation ratio concept, we can see that after 2000, 
the participation of different modes to the network has increased and tends to the market mode of the system. 
Although no important change in the total market share of trading occurs, 
through the passage of time, the contribution of some countries in the network structure has increased. 
The technique proposed in the paper can be useful for analyzing different types of interaction networks 
between countries.

\end{abstract}


\end{frontmatter}



\section{Introduction}
\label{introduction} Since the recent global financial crisis, cross-border
lending and financial contagions have gained importance. This importance
stems from the propagated effects~\cite{Iori2015,Haldane2011} of financial
crises on political and economic situations~\cite{Reinhart2009,Contreras2014}. This fact has prompted a lot of research on the systemic dependence of the
international banking sector ~\cite{Cont2010,Habibnia2017,BattistonH2012,Betz2016,Battiston2017flowofrisk,Battiston2013,Hedayatifar2020}.

One of the most recent approaches for analyzing this situation comes from the notion of complexity~\cite{Cont2010,Jafari2011}. The purpose of
complexity science in finance focuses on the analysis of the structure and
the dynamics of entangled systems. Many scholars have applied complexity
techniques for analyzing financial contagion ~\cite{Glasserman2016,Battiston2013,Battiston2017flowofrisk,
Habibnia2017,Bardoscia2017}. Their findings suggested that connectivity of
financial institutions is the source of potential contagions.

\textit{Random Matrix Theory}  is one of the useful methods for analyzing
the behavior of complex systems~\cite{POTTERS2019,Jiang2014,NamakiNetwork2011,MacMahon2015,Sandoval2012,Jafari2011,Shirazi_2009,Jurczyk2017,Kwapie2012,NamakiStructure2011}. This theory was developed by researchers to describe the situation of
energy levels of quantum systems~\cite{MEHTA1991,MEHTA2004}.

The universality regime of the eigenvalue statistics is the success factor
of Random Matrix Theory~\cite{Plerou2002,LALOUX2000,NamakiRMTtehran2011}.
Based on previous studies, it is shown that when the size of the matrix is
very large, the eigenvalue distribution tends towards a specific
distribution~\cite{NamakiRMTtehran2011}.

Random Matrix Theory has been applied to analyze the behavior of coupling
matrices~\cite{Jafari2011}. This technique divides the contents of the
coupling matrix into noise and information parts. The noise part of the
coupling matrix conforms to the Random Matrix Theory findings and the
information part deviates from them. This concept stems from the idea of
solving the problem of non-stationary cross correlation and measurement
noise, as a result of market conditions and the finite length of time series~\cite{NamakiRMTtehran2011,Plerou2002}.

It is shown that the majority of their eigenvalues agree with the random
matrix predictions, but the largest eigenvalue has deviations from those
estimations~\cite{LALOUX2000,Plerou2002,Wang2013,Kwapie2012}. In essence,
this eigenvalue develops an energy gap that separates it from the other
eigenvalues~\cite{NamakiNetwork2011}. The largest eigenvalue is related to a
strongly delocalized eigenvector that presents the collective evolution of
the system, and this is called the market mode. From this perspective, the
largest eigenvalue's magnitude reflects the coupling strength of the system~\cite{NamakiNetwork2011}.

One of the systems which can be analyzed by the complexity approach, is the
global banking network~\cite{Reyes2011}. In this paper, by applying Random
Matrix Theory as a useful technique from complexity science, we want to
analyze the global banking network.

Our paper is organized as follows. In Section 2 we present our methods and,
in section 3 we apply Random Matrix Theory on the global banking network and
present our findings. Then, in section 4 we conclude. 

\section{Methods}
Primarily Random Matrix Theory has been presented by some scholars in
nuclear physics such as Mehta~\cite{MEHTA1991,MEHTA2004}, for analyzing the
energy levels of complex quantum systems. Subsequently, the mentioned method
helped to address specific issues in other fields, such as finance~\cite{NamakiNetwork2011,NamakiRMTtehran2011,LALOUX2000,Plerou2002}.

Based on the perception from random matrix theory, the eigenvalues --in the
real matrix-- which deviate from the range of the eigenvalues --in the
random matrix-- possess relatively more complete information from the system~\cite{LALOUX2000,NamakiStructure2011,NamakiRMTtehran2011}.

\begin{figure}[tp]
\centering
\includegraphics[width=0.23\textwidth]{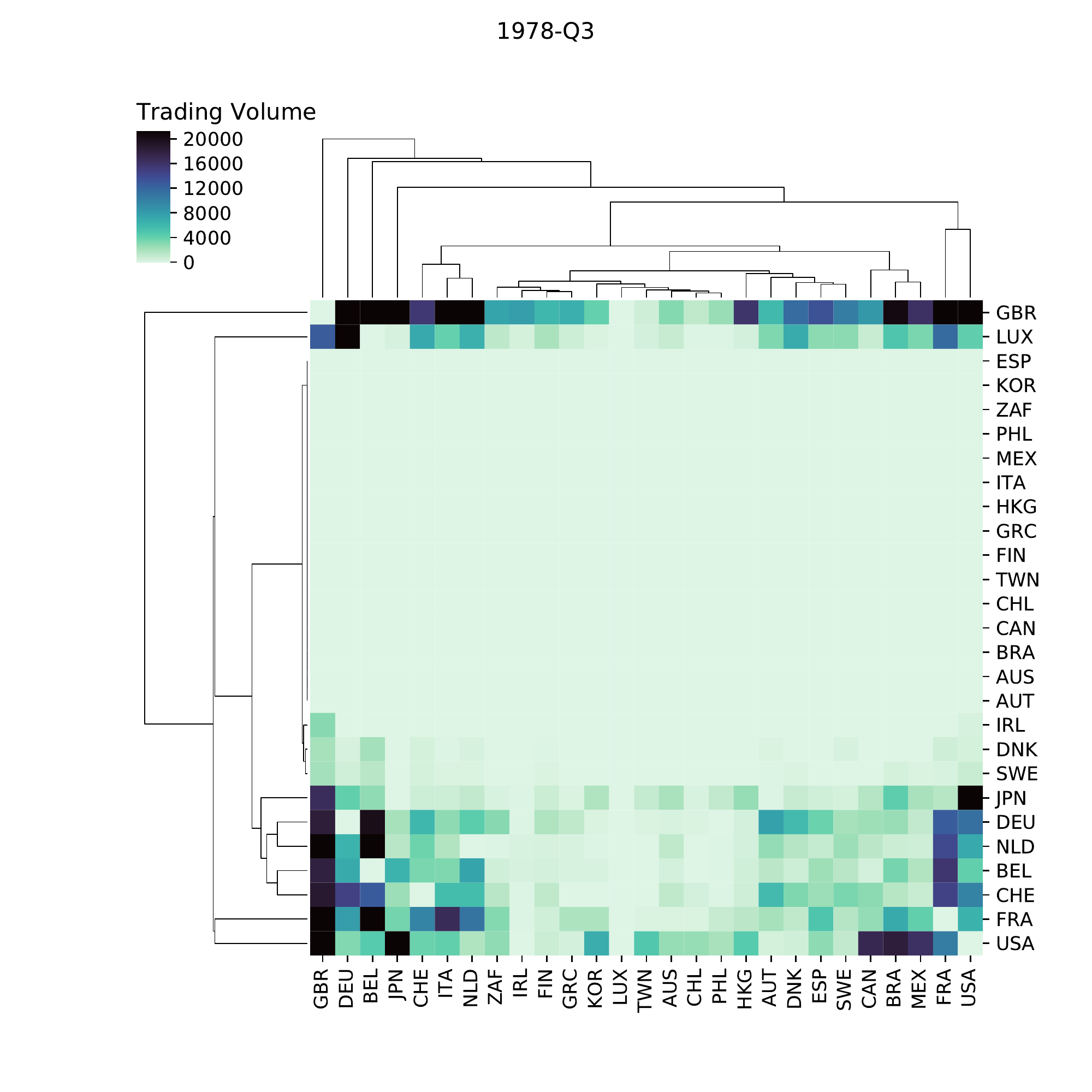}%
\includegraphics[width=0.23\textwidth]{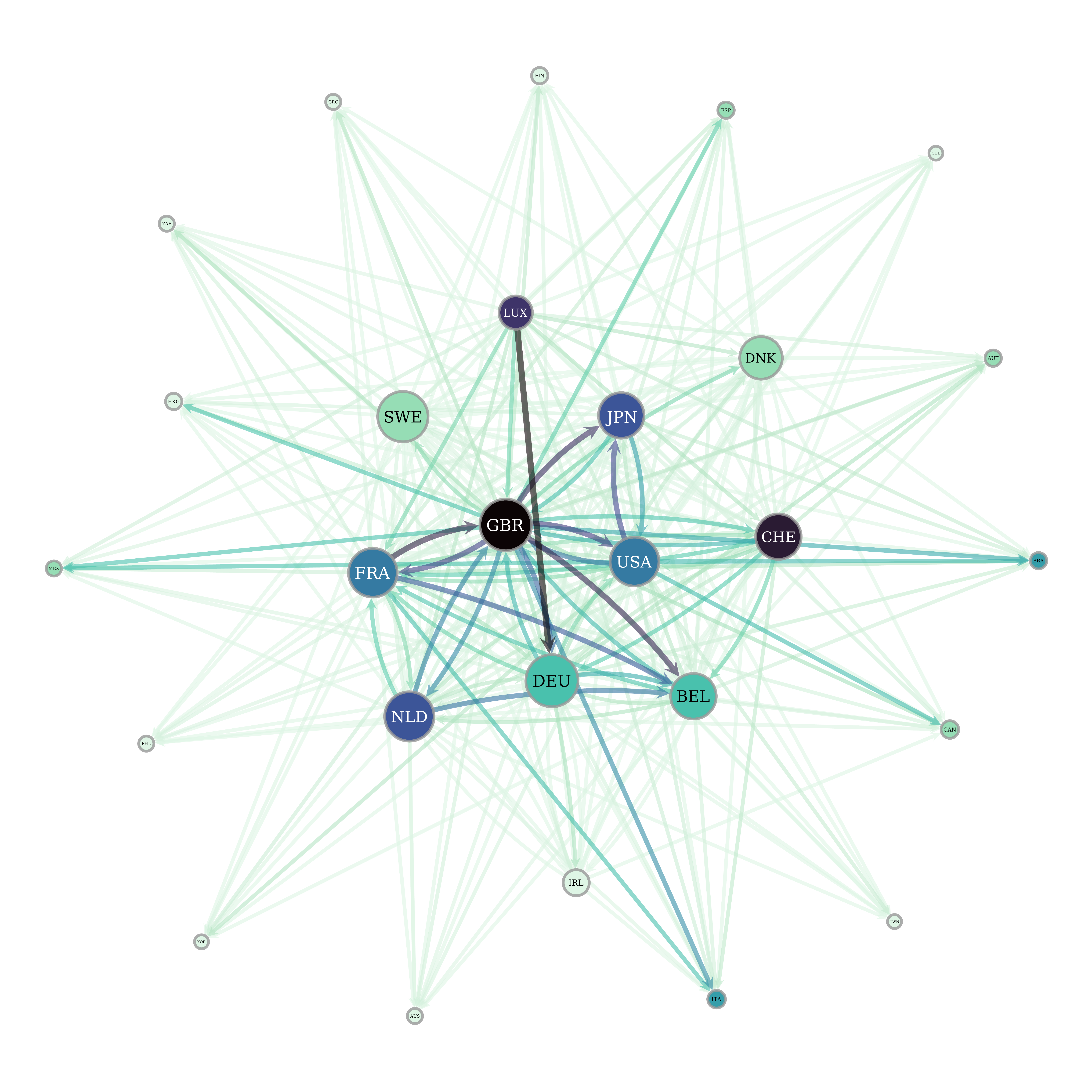}
\par
\includegraphics[width=0.23\textwidth]{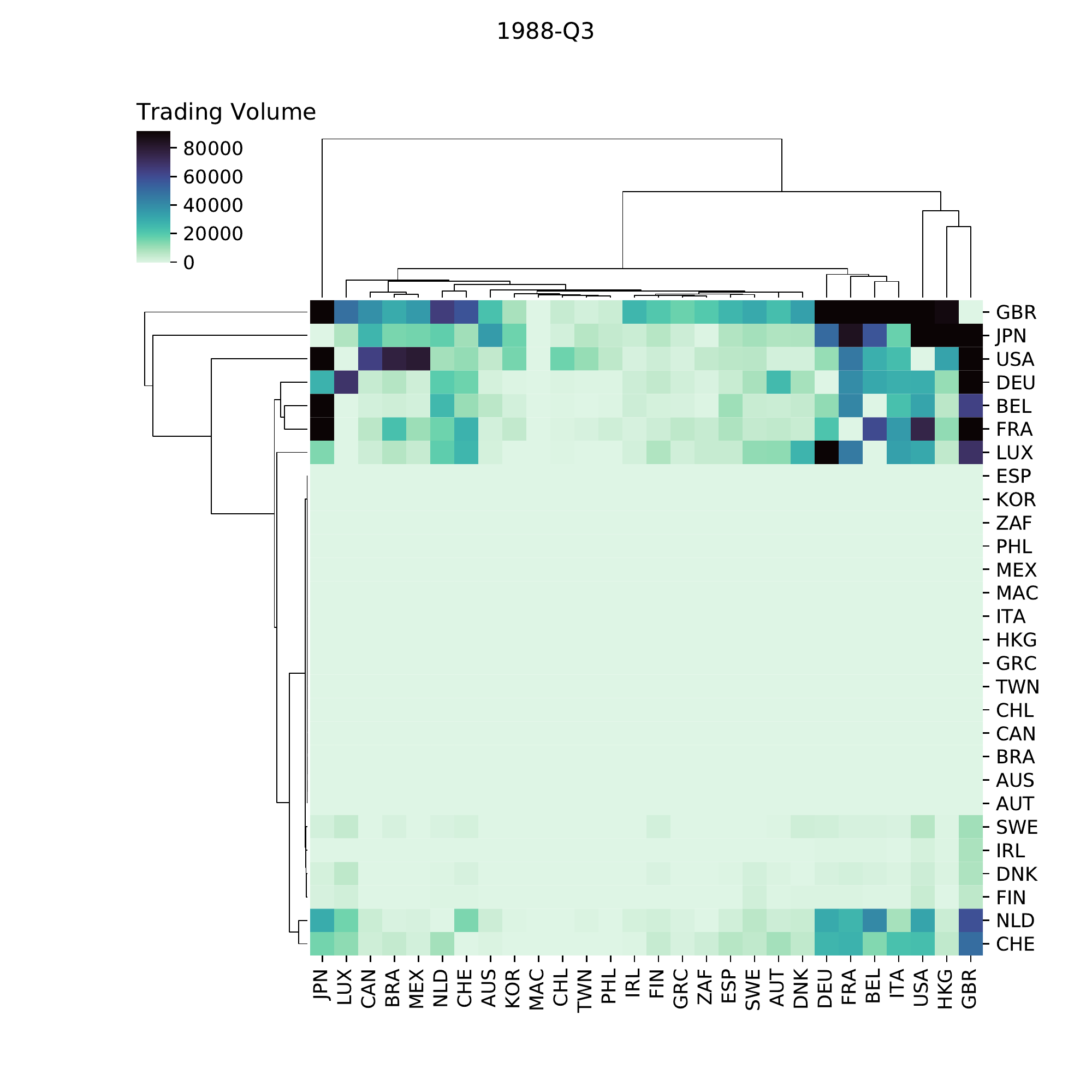}%
\includegraphics[width=0.23\textwidth]{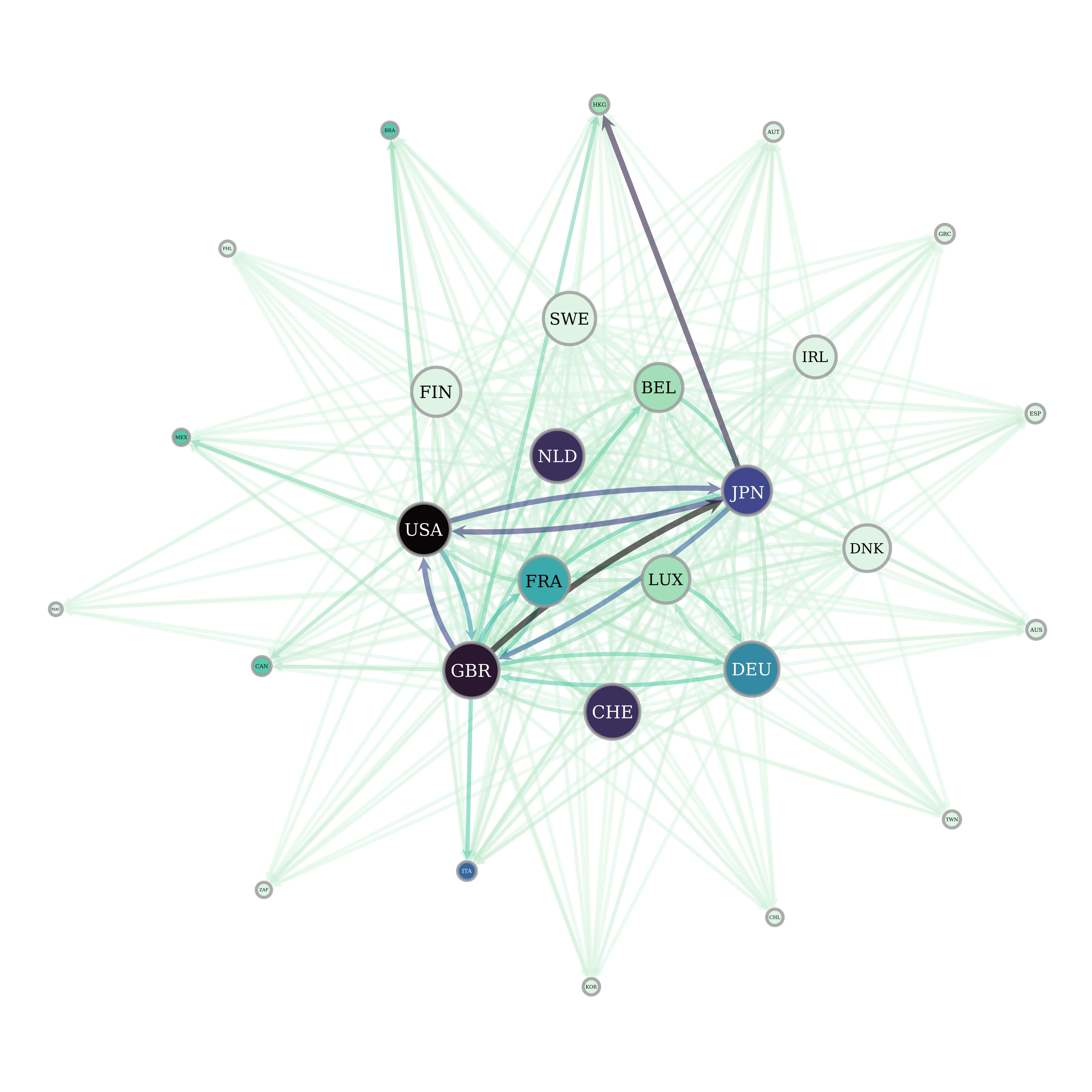}
\par
\includegraphics[width=0.23\textwidth]{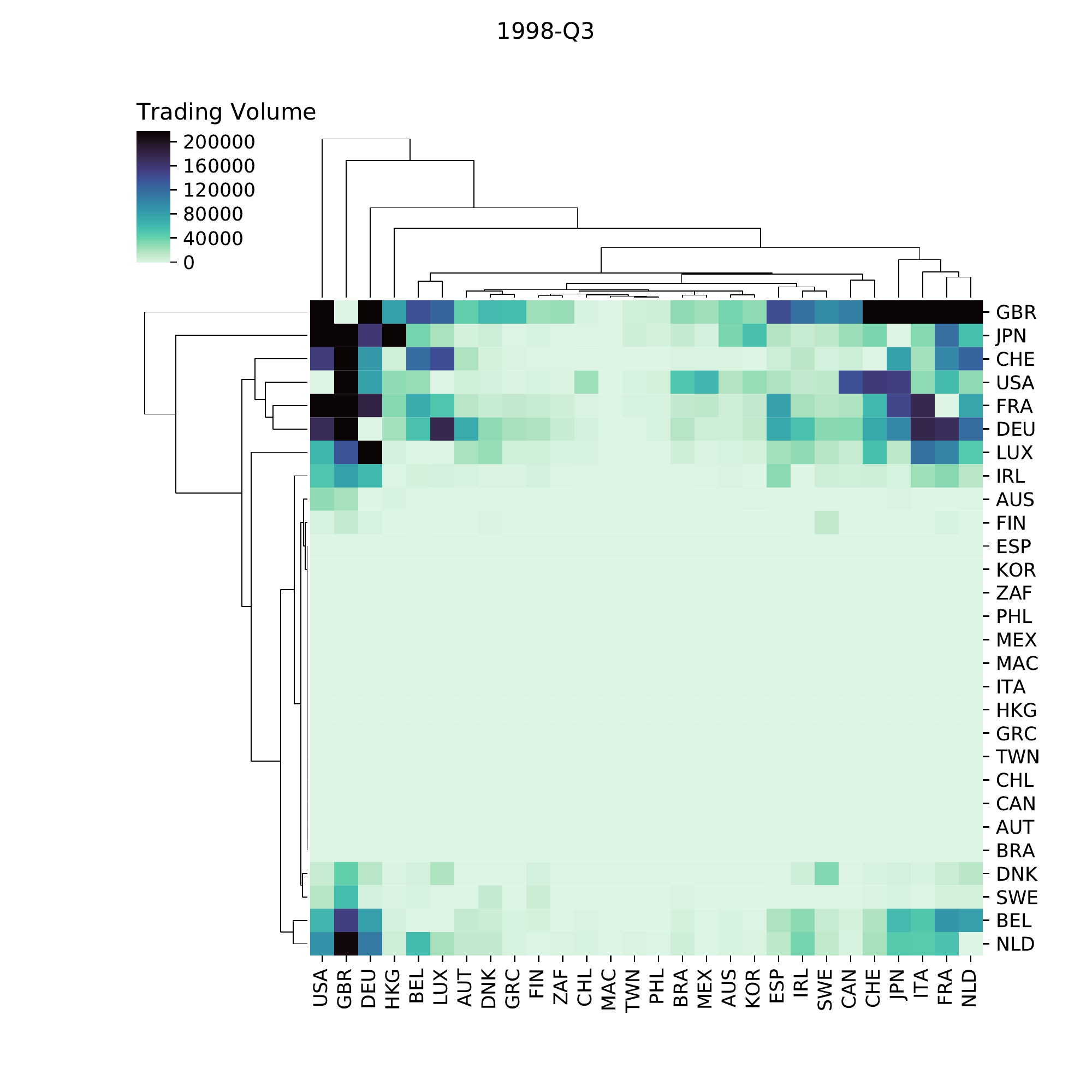}%
\includegraphics[width=0.23\textwidth]{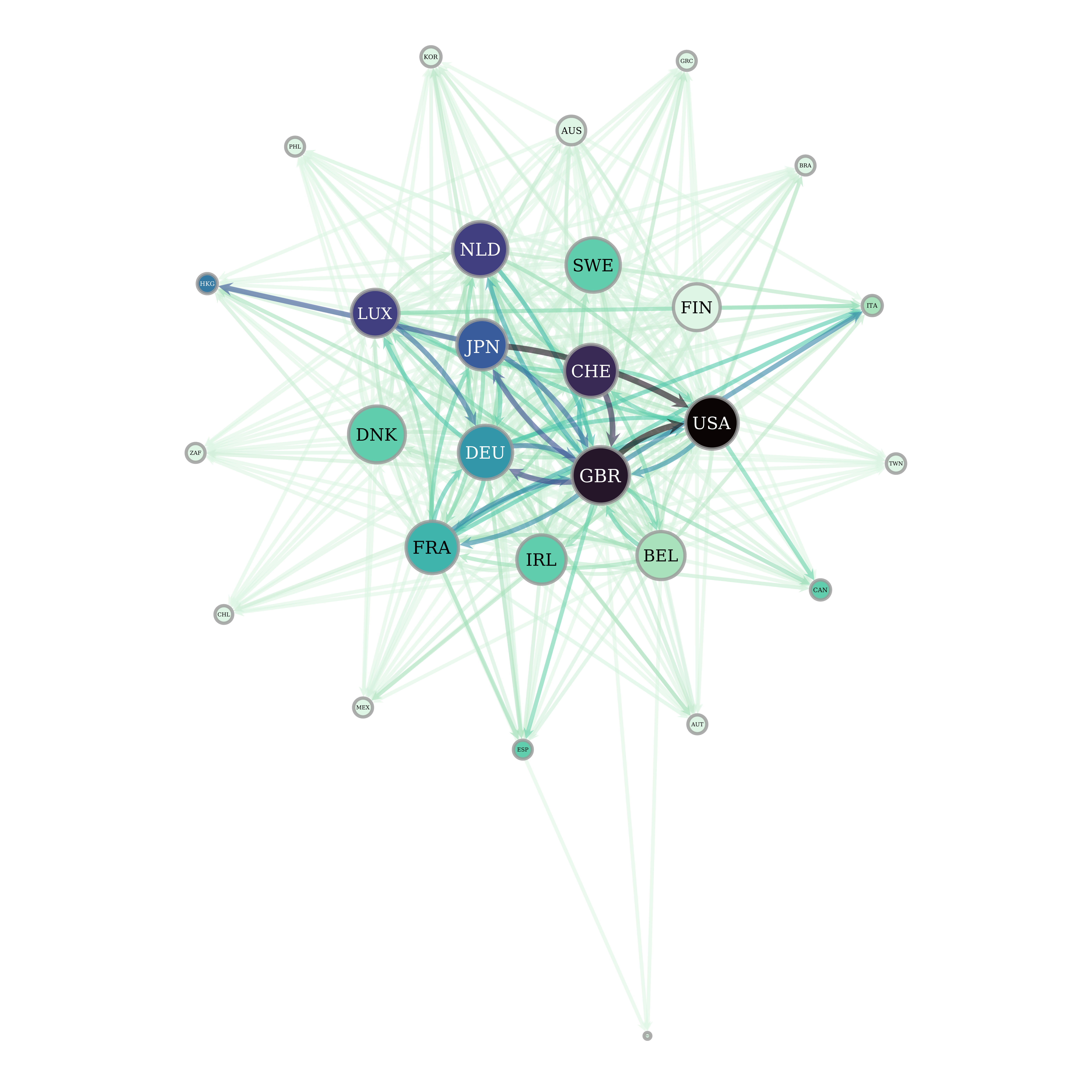}
\par
\includegraphics[width=0.23\textwidth]{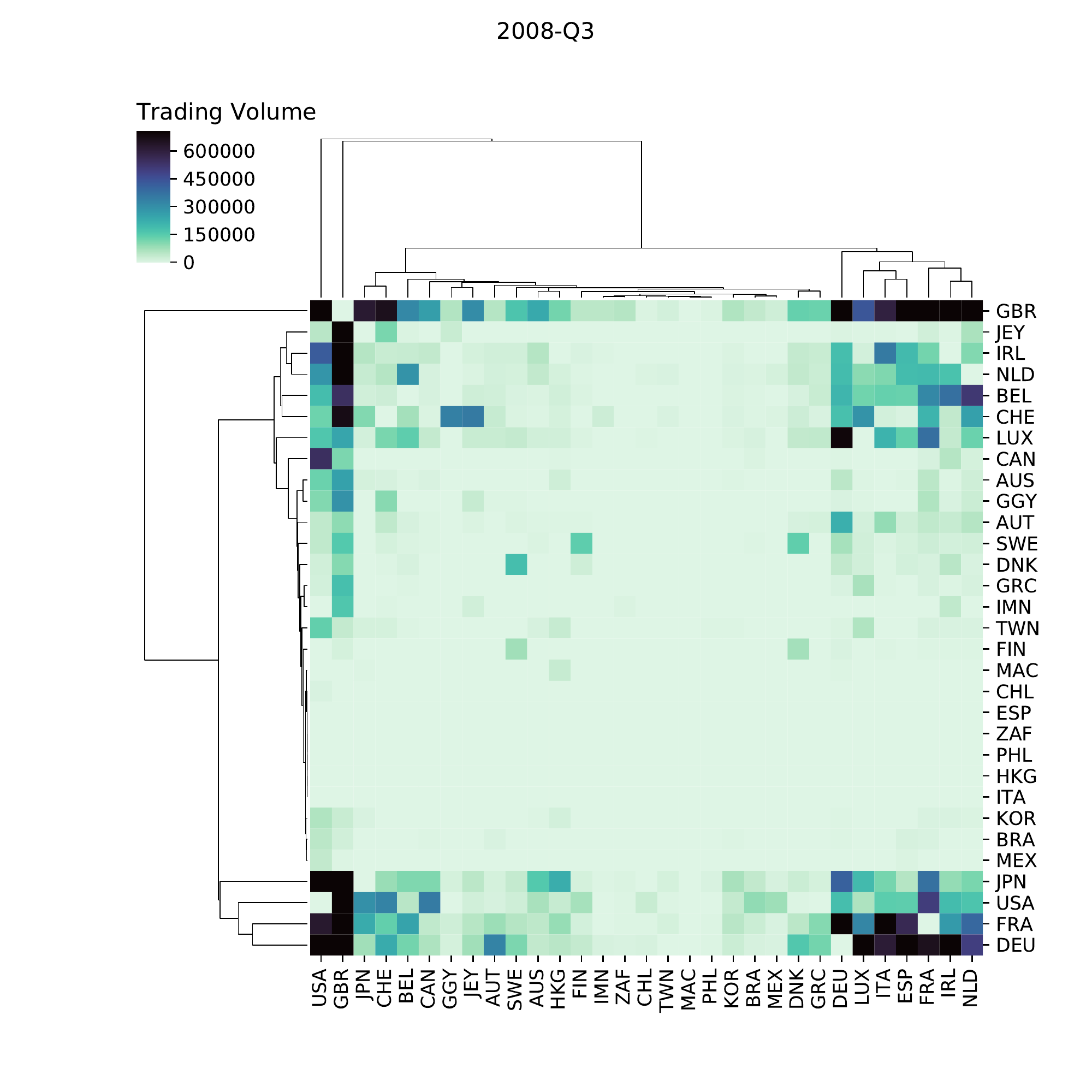}%
\includegraphics[width=0.23\textwidth]{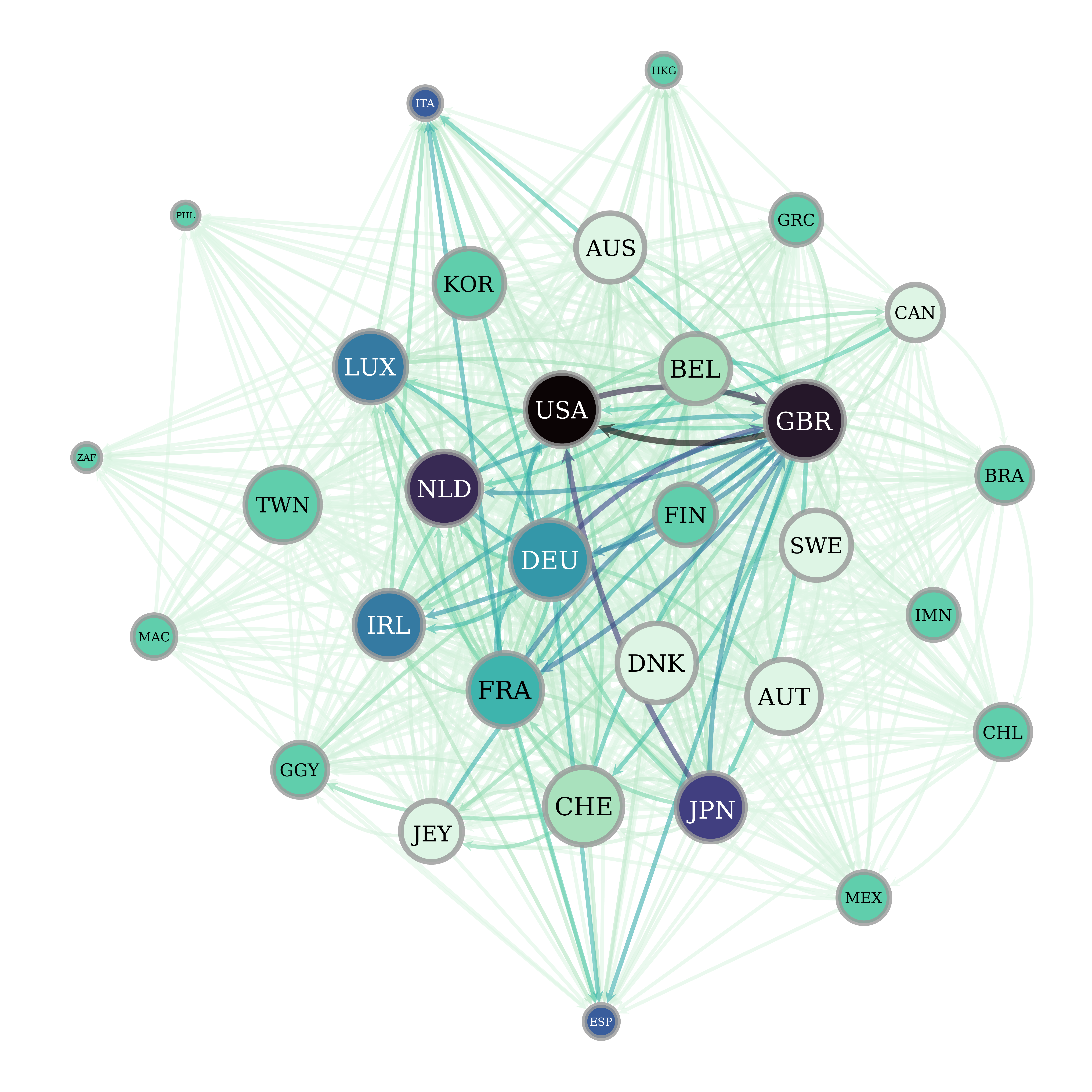}
\par
\includegraphics[width=0.23\textwidth]{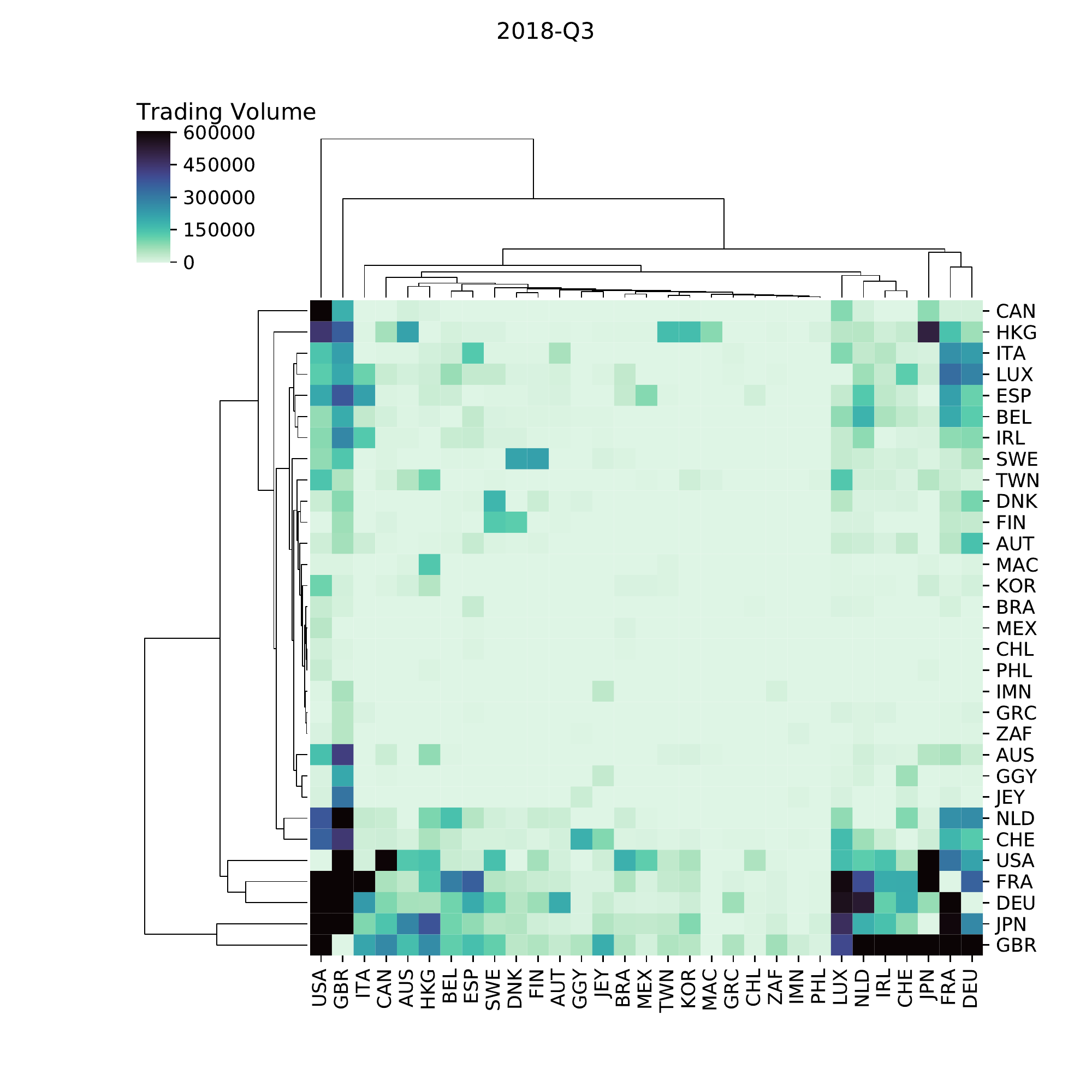}%
\includegraphics[width=0.23\textwidth]{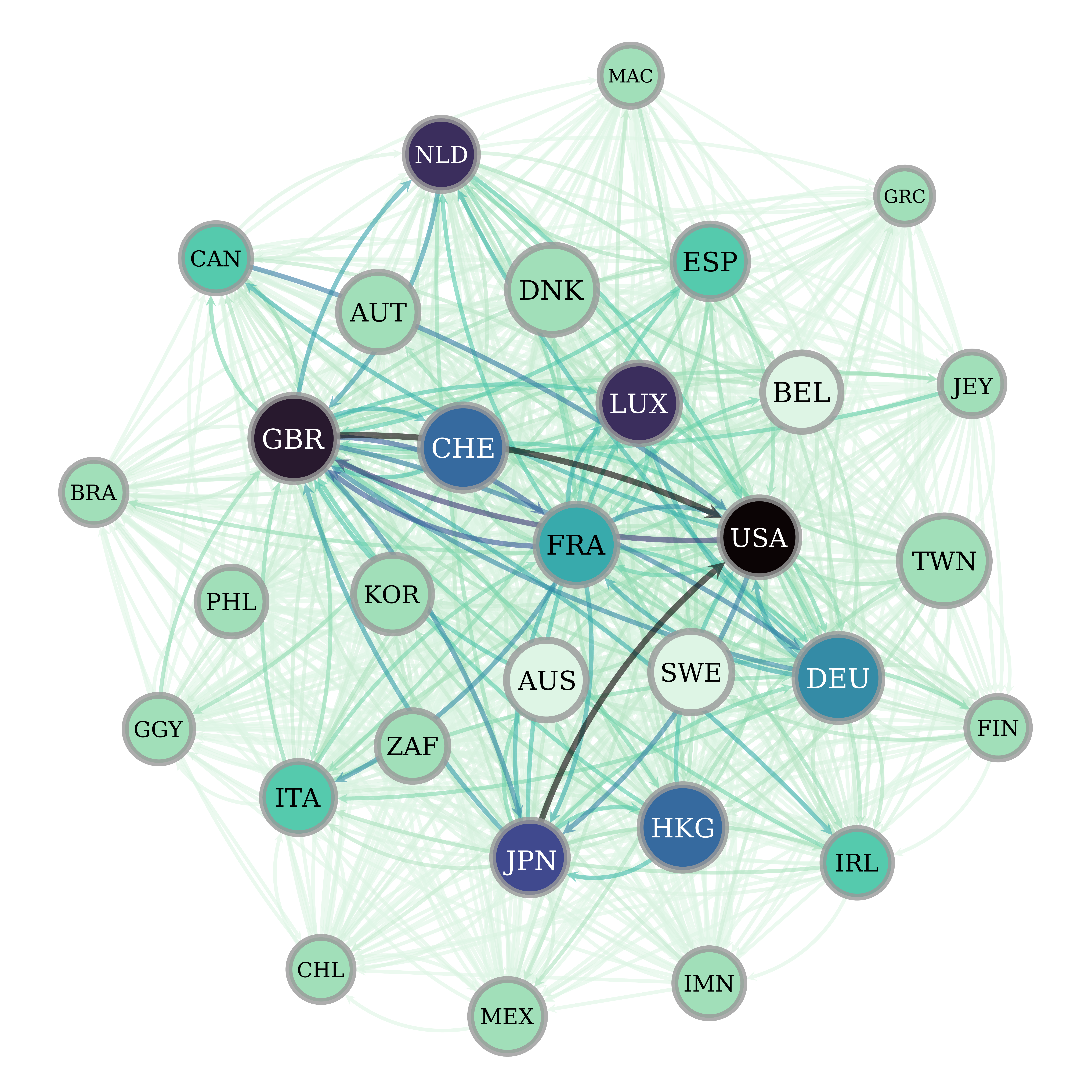}
\caption{The evolution of global banking network is demonstrated for the 5
snapshots of 1978-Q3, 1988-Q3, 1998-Q3, 2008-Q3 and 2018-Q3. Left column;
shows the evolution of trading matrices between countries. In order to
extract the structure of their communities, we have applied the dendrogram
weighted matrices. Right column; shows the evolution of network topology.}
\label{network_evolution}
\end{figure}
\begin{figure}[t]
\centering
\includegraphics[width=0.4\textwidth]{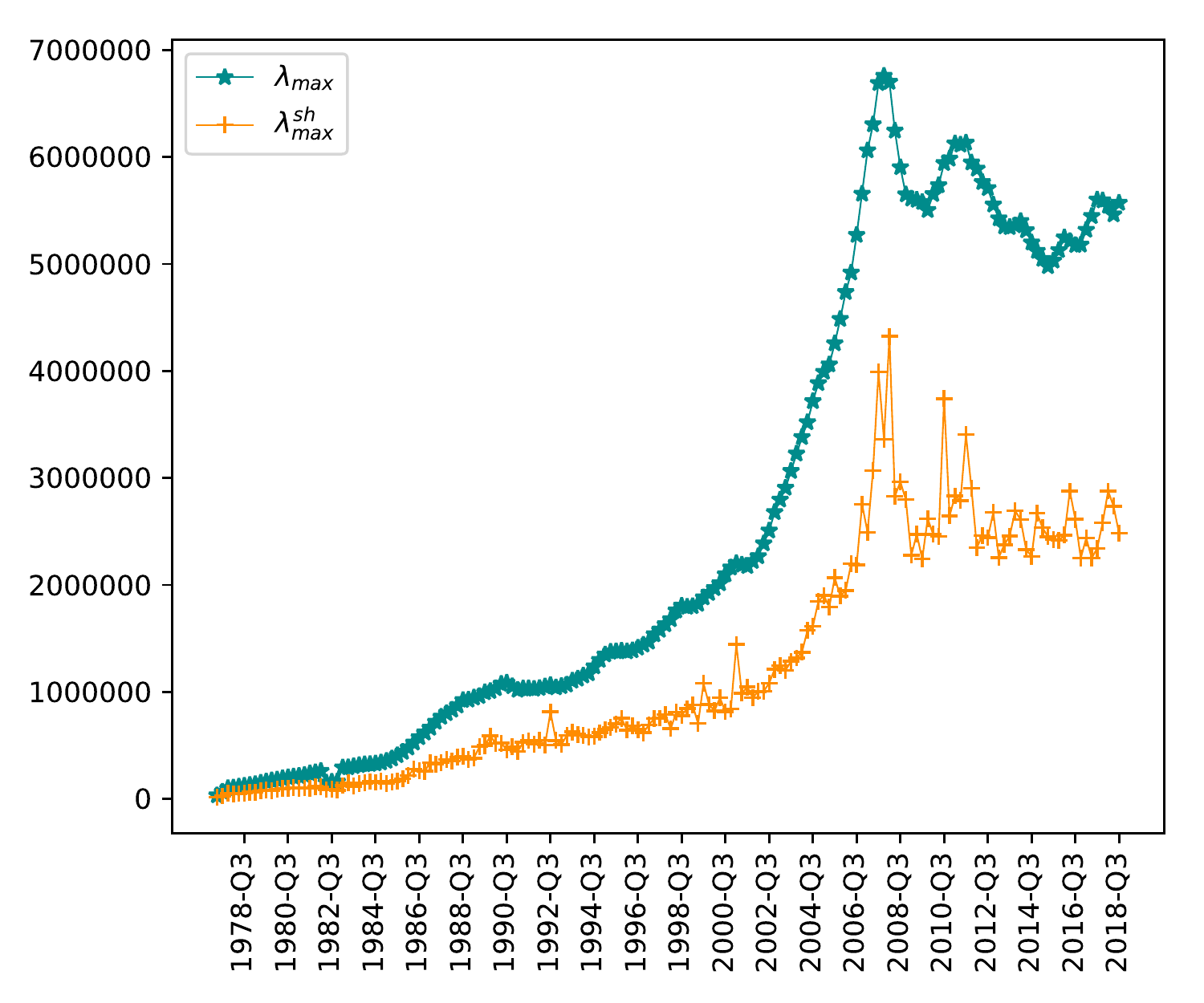}  %
\includegraphics[width=0.4\textwidth]{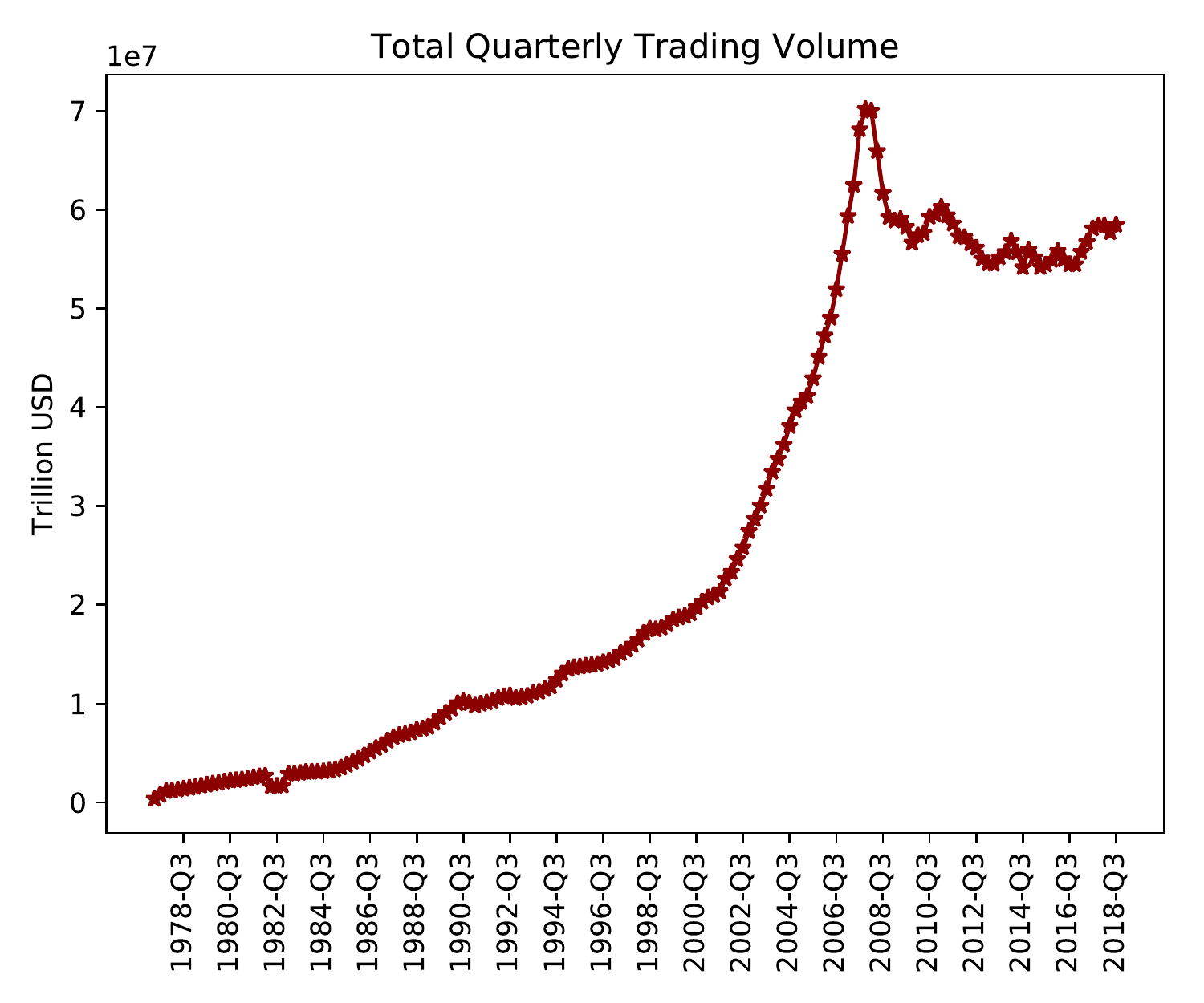}
\caption{up) The evolution of the largest eigenvalue, $\lambda_{max}$, of the global banking network and its shuffled, $\lambda_{max}^{sh}$, are depicted. down) The evolution of total trading volume is
demonstrated.}
\label{lambdamax}
\end{figure}
\begin{figure}[tp]
\centering
\includegraphics[width=0.4\textwidth]{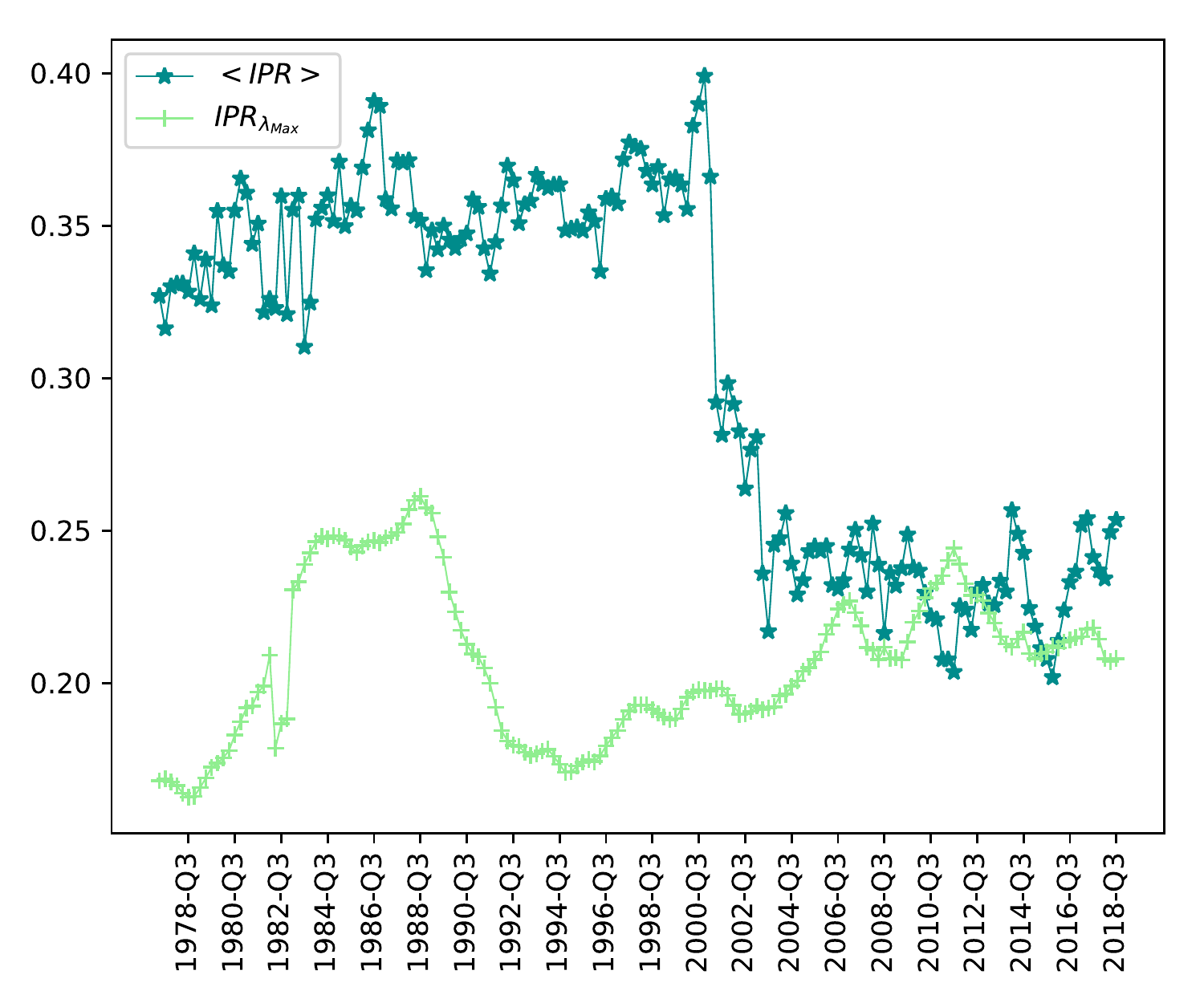} 
\caption{It is depicted that overtime $<IPR>$ has tended to $IPR_{\lambda _{max}}$. It implies that the contribution of countries has
generally increased.}
\label{IPR}
\end{figure}
\begin{figure*}[pht]
\centering
\includegraphics[width=1\textwidth]{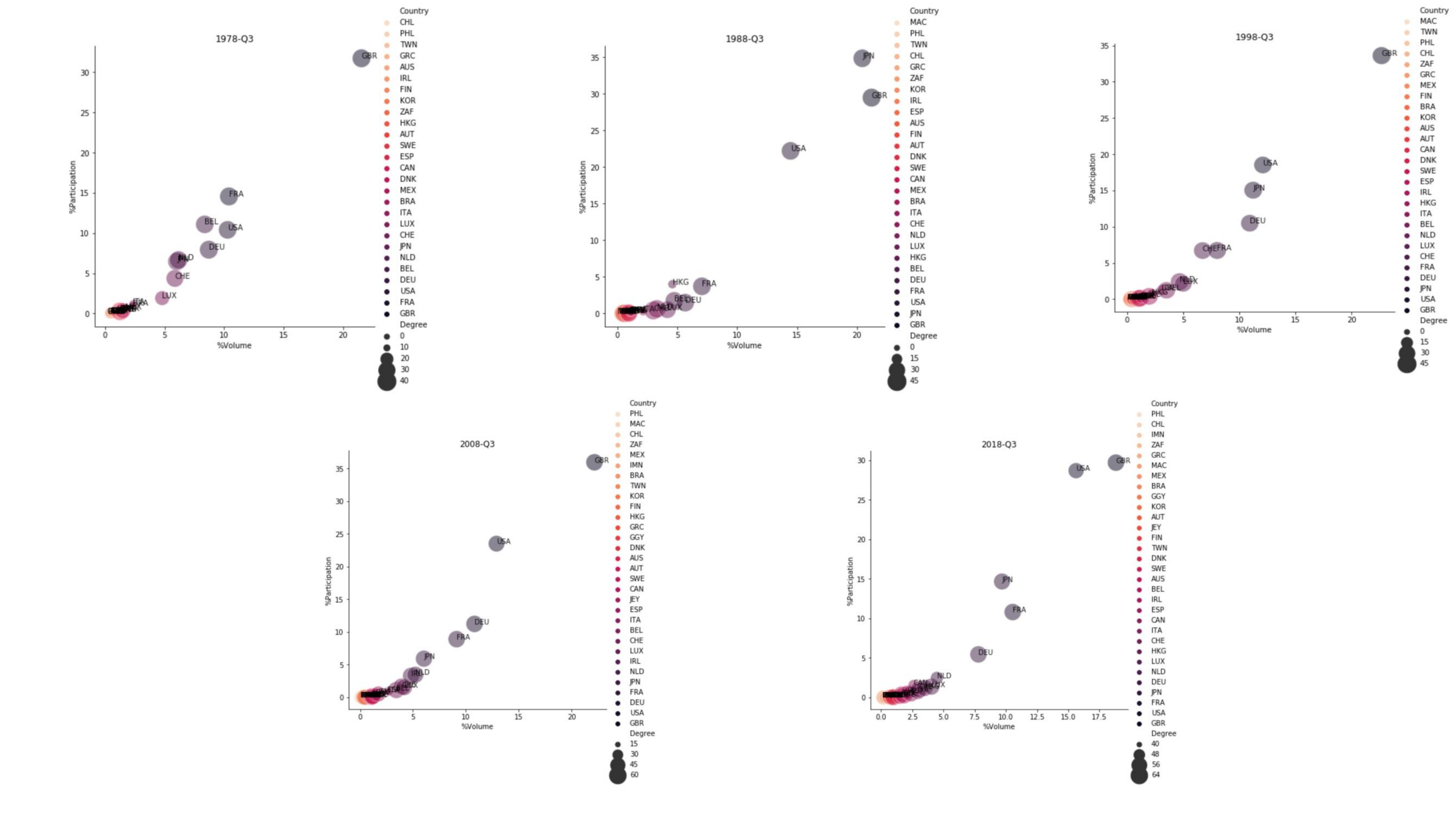}
\caption{shows the \% participation of each country in the eigenvector --
corresponding to the largest eigenvalue -- versus \%($\frac{volume_{j}}{\Sigma _{i}^{N}Volume_{i}}$.)}
\label{ParticipationVol}
\end{figure*}
In Random Matrix Theory, there is a parameter named as the \textit{Inverse Participation Ratio} ${IPR}$ which is based on the theory of Anderson's localization~\cite{Lim2009}, and it computes the number of components which significantly
participate in each eigenvector. This notion shows the effect of components
of each eigenvector, and specifically how the largest eigenvalues deviate
from the bulk region which is densely occupied by eigenvalues of the random
matrix. Based on the previous papers~\cite{NamakiNetwork2011,Saeedian2019}, $IPR$ can be applied as an indicator for measuring the collective behavior of
the networks. The formula of this concept is as follows: 
\begin{equation}
IPR(k)=\frac{1}{\sum_{l=1}^{n}(u_{l}^{k})^{4}};  \label{eq_IPR}
\end{equation}%
where $l=1,\ldots ,n$ and $u_{l}^{k}$ is the $l^{th}$ element of $k^{th}$
eigenvector ($lk$). To further clarify the concept, one may consider
examples below:

\indent-- In case all elements of a certain eigenvector are equal to $\frac{1}{\sqrt{N}}$, $IPR$ will be equal to $N$. This implies that whole elements
are significantly influential on the systems' behavior.

\indent -- On the other hand, if just a single element is equal to $1$ and
the others are equal to $0$, $IPR$ would be equal to $1$. This implies that
only this component is effective in the corresponding eigenvector.

Hence, one can perceive that $IPR$ clarifies the number of influential
elements in a certain eigenvector.

\section{Analysis of Global Banking Network by Random Matrix Theory}

The banking industry is one of the most important sectors in finance. In
this regard, one of the significant aspects of financial contagion is the
emergence and transmission of crisis throughout the banking network. In Fig.~\ref{network_evolution}, the evolution of the global banking network in 5 snapshots (1978-Q3, 1988-Q3, 1998-Q3, 2008-Q3 and 2018-Q3) has been
depicted. The left panel in Fig.~\ref{network_evolution} shows the
dendrogram structure of communities for trading weighted matrices. Also, the
right column shows the evolution of the network topology. As depicted, the
network has been denser over time. Not only the contributions have risen,
but also the peripheral nodes are arranged closer and connected to the
central section.

In this study, we apply Random Matrix Theory for the data of BIS bilateral
locational statistics provided by the \textit{Bank for International
Settlements (BIS)}~\cite{BIS} from 1978 until 2019. This data includes all
`core' countries (the qualifier `core' is used by many researchers such as~\cite{Reyes2011}, for 31 countries which regularly report their financial
data to BIS).

We create a weighted and directed financial transaction network
corresponding to each quarter from 1978 until 2019. Each link corresponds to
a loan given by a certain country to another one. Previous studies
specifically shed light onto countries' dependency network and showed an
increase in the dependency structure of the network of those countries
during the passage of time~\cite{Reyes2011}. As already discussed, Random
Matrix Theory is a powerful approach for analyzing complex systems. In this
paper we apply this concept for the analysis of the global banking network
as a complex network. For this purpose, we choose the shuffling technique
for the construction of a random matrix. The shuffling method which is applied in this research is randomization of bilateral trading volume (or links) in the network. It means that the PDF remains unchanged and the bilateral trading relations will be shuffled. The Shuffled matrix is an indication of no information in the system.

The global banking network possesses an adjacency matrix. This matrix can
intrinsically be explained by the eigenvalue decomposition methods~\cite{barucca2019eigenvalue}. The eigenvector corresponding to the largest
eigenvalue, $\lambda_{max}$, is the most significant and is the market mode
of the network~\cite{Utsugi2004,NamakiNetwork2011,LALOUX2000,Plerou2002}.

In this regard, we assess the temporal behavior of the largest eigenvalue,
as shown in Fig.~\ref{lambdamax}.

By evaluating the behavior of $\lambda _{max}$ and comparing it with the $\lambda _{max}$ of the corresponding shuffled matrix in Fig.~\ref{lambdamax}, one can observe the information content of the market mode. As depicted in
Fig.~\ref{lambdamax}, the temporal behavior of the largest eigenvalue in the
banking interaction matrix, is totally different from that of the largest
eigenvalue in the shuffled matrix. This phenomenon determines the existence
of information content embedded in the largest eigenvalue of the banking
interaction matrix.

When it comes to Fig.~\ref{lambdamax}, the behavior of the maximum
eigenvalue has been ascending. This issue~-- as stated
before--- has bilateral effects.

The reasoning behind this is that on the one side, it causes more strength
and stability in the network, whilst on the other side, it yields to a more
agile contagion throughout the network~\cite{BattistonH2012}. In the
post-crisis era after 2008, simultaneous to a decrease in the maximum
eigenvalue, the collective behavior of the system has reduced and
accordingly, local identities have been more significant.

Since the so-obtained eigenvalue does not describe all the details and
properties of the collective behavior, one should investigate other
quantities in the network.

It is observable that during the global financial crisis, a structural
emergence with an increase in the difference between $\lambda_{max}$ and $\lambda _{max}^{sh}$, Fig.~\ref{lambdamax}, has occurred.

However, after the crisis, a significant decrease in the behavior of the
largest eigenvalue of the banking matrix relating to that of the shuffled
matrix has emerged.

Based on the above concepts, one of the best approaches for
analyzing the global banking network is the Random Matrix Theory technique.


As already discussed, one should keep in mind that $IPR$ possesses the
ability of information extraction from the collective behaviors of the
systems.

In Fig.~\ref{IPR}, By comparing $<IPR>$ and $IPR_{\lambda_{max}}$, one is
able to distinguish the temporal evolution of participation in the network.

In Fig.~\ref{IPR}, we investigate the \textit{inverse participation ratio
(IPR)} in a temporal process. In this context, by focusing on the
 mean inverse participation ratio , $<IPR>
$, and also, the inverse participation ratio of the
largest eigenvalue corresponding to the largest
eigenvector , we investigate banking behaviors of the
countries and their influences on the network structure and the market
trend. In Fig.~\ref{IPR}, $IPR_{mean}$ implies the effectiveness of the
banking system of most countries on the global network. However, from the
temporal behavior of $IPR_{\lambda _{max}}$, we observe that over time, less
participation from those countries on the largest eigenvector emerges.

In Fig.~\ref{ParticipationVol}, $\%Participation$ stands for the
contribution percentage of each country in the eigenvector corresponding to
the largest eigenvalue. $\%Volume$ is the trading volume of a
country divided by the total trading volume.
Hence, $\%Participation$ shows the contribution in the structure, and, $%
\%Volume$ shows the contribution in the total trading volume. Thereby, Fig.~\ref{ParticipationVol} visualizes the contributions in the structure versus
the contribution in trading volume within each year. In 2018-Q3, for the US,
while the percentage of contribution in the structure has been approximately
constant, the percentage of contribution in trading volume decreased.

\section{Conclusion}

In this paper, by applying Random Matrix Theory, the global banking network
is investigated. For this purpose, we compute the matrix of interaction of
banking sectors of BIS countries, and then by using the Random Matrix Theory
approach, the behavior of the largest eigenvalue and Inverse Participation
Ratio of this eigenvalue, as the market mode of the system over time, has
been analyzed. The value of the largest eigenvalue increases during the
passage of time. By observing the behavior of trading volume, it is shown
that these increases stem from the expansion of the network to some extent.
Also, by comparing with the shuffled one, we can deduce that the system gets
a specific structure. Generally speaking, the global banking network, today,
is more dense and interconnected. Also, we can see that after the year 2000
the value of the mean $IPR$ has dropped and converged to $IPR_{\lambda_{max}}$. It means that more countries have become more influential on the
global banking network. Furthermore, despite small changes in the share of
total trading volume, some countries such as the UK, have become more
important in the network structure.

As a concluding remark, the identities of banking systems of BIS countries
stems from two parts, i.e. i) from their own identities individually and,
ii) from their interactions in the global banking network. As a suggestion
for further work, one can construct the interaction matrices of the
countries based on other variables such as commercial interactions and so
on.


\bibliographystyle{elsarticle-num}
\bibliography{GlobalBanking.bib}



\end{document}